\documentclass[submit]{elex2024}
\usepackage{hyperref}
\usepackage{cite}
\usepackage{algorithmic}
\usepackage{graphicx}
\usepackage{textcomp}
\usepackage{xcolor}
\usepackage{booktabs}  
\usepackage{threeparttable} 
\usepackage{multirow} 
\usepackage{array} 
\usepackage{makecell} 
\usepackage{graphicx} 
\usepackage{siunitx} 

\type{LETTER}

\title{AES-RV: Hardware-Efficient RISC-V Accelerator with Low-Latency AES Instruction Extension for IoT Security}

\author{Van Tinh Nguyen}{1}<0000-0001-7583-8631>
\author{Phuc Hung Pham}{1}
\author{Vu Trung Duong Le}{2}[le.duong@naist.ac.jp]<0000-0002-0438-3809>
\author{Hoai Luan Pham}{2}
\author{Tuan Hai Vu}{2}
\author{Thi Diem Tran}{3}

\affiliate{1}{Le Quy Don Technical University, 236 Hoang Quoc Viet Street, Bac Tu Liem District, Hanoi, Vietnam}
\affiliate{2}{Nara Institute of Science and Technology, 8916–5 Takayama-cho, Ikoma, Nara, 630-0192 Japan}
\affiliate{3}{University of Information Technology, Vietnam National University, Ho Chi Minh City, 700000, Vietnam}

\vol{21}
\no{1}

\doi{10.1587/elex.XX.XXXXXXXX}
\received{2024/10/10}
\accepted{2024/10/10}
\publicized{2024/12/31}
\copyedited{2024/12/31}

\begin{document}

\maketitle 

\begin{abstract}
The Advanced Encryption Standard (AES) is a fundamental cryptographic algorithm widely used to secure data in embedded systems, IoT devices, and cloud computing platforms. However, recent research on AES hardware accelerators face challenges in achieving high performance and hardware efficiency, particularly when supporting multiple modes and key sizes. To address these limitations, this paper proposes a hardware-efficient RISC-V accelerator with low-latency AES instruction extension (AES-RV), designed to enhance both processing speed and energy efficiency across various AES configurations. Specifically, AES-RV incorporates three key optimizations: high-bandwidth internal buffers for continuous data processing, a specialized AES unit with low-latency custom instructions, and system pipelining with a ping-pong memory transfer mechanism. The AES-RV accelerator is implemented and evaluated on a real-time Xilinx ZCU102 FPGA system-on-chip (SoC), utilizing 29,608 FFs, 32,483 LUTs, and 12 BRAMs. Performance comparisons against a baseline RISC-V implementation for multiple AES modes and key sizes demonstrate latency improvements ranging from 195.5 times to 255.97 times. Additionally, evaluations against powerful CPUs and GPUs in real-time AES executions reveal energy efficiency gains of 9.92 times to 453.04 times. Compared to state-of-the-art AES hardware accelerators, AES-RV achieves throughput improvements of 13.56 times to 33.52 times, energy efficiency enhancements of 2.36 times to 58.76 times, and area efficiency gains of 91.42 times to 638.8 times.
\end{abstract}



\section{Introduction}

    RISC-V, introduced by the Berkeley research group in the late 2010s, is an open-source CPU architecture designed to promote flexibility and innovation. Its open Instruction Set Architecture (ISA) enables developers to design custom processors without licensing fees, facilitating advancements in specialized hardware. With high compatibility across various applications and superior energy efficiency, RISC-V is an ideal choice for resource-constrained devices \cite{sharma2022survey, a1, a2,a3,a4,a5}. Specifically, RISC-V accelerates encryption for IoT devices, supporting the implementation of traditional security algorithms such as SM3, SM4, and SHA-256 \cite{Le2024RISCVCoprocessor}. These applications ensure data security and enhance performance in IoT systems. Although RISC-V offers numerous advantages, its ISA continues to evolve to meet emerging demands in fields like artificial intelligence and cloud computing.
	
    The Advanced Encryption Standard (AES) \cite{Le2024RISCVCoprocessor,a6,a7,a8,a9}, standardized by the National Institute of Standards and Technology (NIST), is fundamental to modern digital security. Its adoption has expanded to cost-sensitive systems such as IoT devices, edge servers, fog servers, personal computers, and smartphones, where it secures data, meetings, video content, and confidential files \cite{a10, a18,a19,a20,a21,a22,a23}. These platforms require AES processing solutions that are energy-efficient to preserve battery life, high-performing to meet server security demands, and flexible to support various AES modes (ECB, CBC, CTR, CFB) and key sizes (AES-128, AES-192, AES-256) \cite{9759828, a11,a12,a13,a14,a15,a16,a17}. Implementing AES processors on the RISC-V architecture addresses these needs by offering customizable, high-speed, and low-cost hardware solutions. RISC-V allows for tailored AES integration, enhancing performance and reducing hardware complexity. Integrating AES capabilities into RISC-V-based systems enables the development of cost-effective, high-performance security solutions tailored to the specific requirements of contemporary information security infrastructures.

    Several studies have explored AES acceleration \cite{le2024rvcp}, which is crucial for securing IoT, edge, and personal computing devices while maintaining efficiency and flexibility \cite{9759828}. In \cite{10848921}, an AES-256 accelerator based on a 5-stage RV32IMFC RISC-V core improved speed by 82–84\% over software solutions. However, it lacked a dedicated AES ISA extension, limiting integration flexibility. Similarly, \cite{9547962} proposed a custom ISA for AES on the IBEX core, achieving up to 662 times higher energy efficiency than TinyAES. Yet, its design prevented parallel execution with the processor, reducing throughput in multi-task IoT applications. Meanwhile, \cite{a10} introduced a RISC-V cryptographic accelerator with dual concatenable 32-bit ALUs, enabling either two parallel 32-bit operations or a combined 64-bit operation. Despite achieving a 1.7–3.0 times processing speedup, it only supported AES-128 and lacked key expansion optimization, limiting its flexibility. Additionally, \cite{le2024rvcp} proposed a flexible cryptographic unit supporting multiple cryptographic functions. However, it did not optimize AES key expansion or support multiple AES modes, leading to reduced throughput and flexibility. Overall, these RISC-V-based AES implementations still struggle to balance flexibility and hardware efficiency, leaving room for further improvements in integration, parallelism, and mode support.

    To overcome current challenges, this paper proposes AES-RV, a hardware-efficient RISC-V accelerator with low-latency AES instruction extensions for enhanced flexibility and energy efficiency. AES-RV integrates three key optimizations: high-bandwidth buffers for continuous data flow, a specialized AES unit with low-latency instructions, and pipelined processing with a ping-pong memory mechanism. Implemented on a Xilinx ZCU102 SoC FPGA, AES-RV outperforms existing AES platforms by supporting all AES modes and key sizes while meeting edge device constraints.

    The remainder of this paper is organized as follows: Section \ref{sec2} presents the details of the AES-RV proposal. Next, Section \ref{sec3} shows the evaluation results of the AES-RV. Finally, section \ref{sec4} concludes the paper.

	\section{Proposed AES-RV Architecture}
	\label{sec2}
	\subsection{System Architecture Overview}
	
	\begin{figure}[t]
		\centering
		\includegraphics[width=0.47 \textwidth]{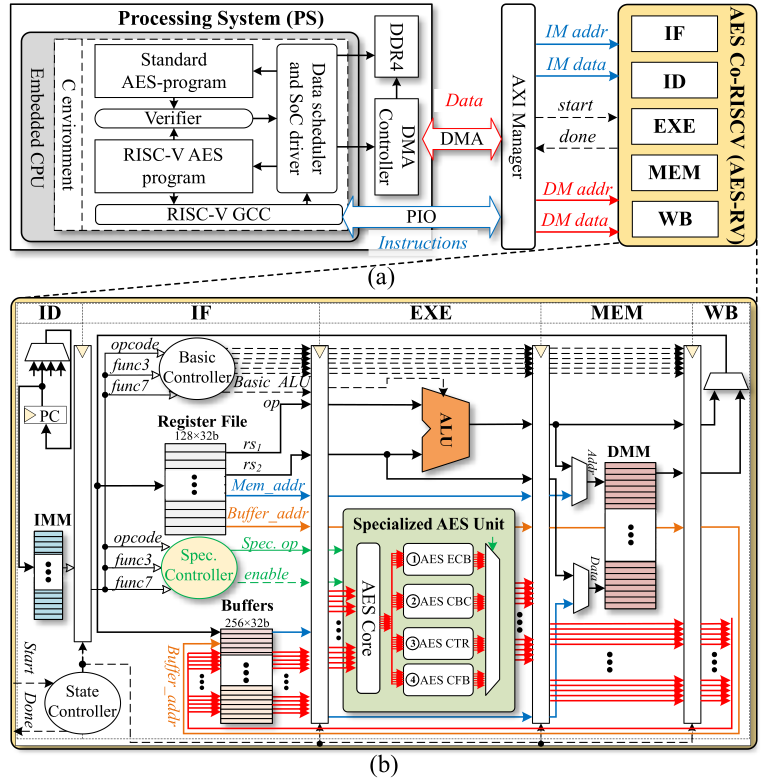}
		\vspace{-2mm}
		\caption{(a) System architecture overview of AES-RV at the SoC. (b) AES-RV architecture.}
		\vspace{-2mm}
		\label{fig:overview}
	\end{figure}

	The overview SoC architecture of the AES Co RISC-V, as illustrated in Fig. \ref{fig:overview}(a), comprises two main components: the Processing System (PS) and the AES-RV module. The PS, controlled by an embedded CPU, executes the C environment to run both standard AES and RISC-V AES programs. The RISC-V AES program is compiled and transferred to the AES-RV for storage in the instruction memory. Meanwhile, the computational data is prepared by the Data Scheduler and SoC driver, facilitating data exchange with the AES-RV’s data memory. To optimize data transfer efficiency, instruction data is transmitted via PIO transfer, while large-scale computational data is transferred through Direct Memory Access (DMA).
	
	The AES-RV module consists of two primary components: the AXI Manager and the AES-RV core. The AXI Manager decodes data exchanged between the PS and the AES-RV. Specifically, computational data is stored in data memory (DM), compiled hexadecimal code is stored in instruction memory (IM), and control signals such as start and done are managed by two controllers within the AES-RV. Similar to conventional RISC-V architectures, the AES-RV core features a five-stage pipeline: instruction fetch (IF), instruction decode (ID), execution (EXE), memory access (MEM), and writeback (WB). However, traditional RISC-V cores utilize a basic ALU with two input sources, performing one operation per instruction, which results in high latency for AES computations. To address this, the AES-RV core, illustrated in Fig. \ref{fig:overview}(b) integrates a Specialized AES Unit (SAU) to accelerate cryptographic operations. The SAU is a highly flexible ALU designed to support multiple AES modes, including ECB, CBC, CTR, and CFB, with various key sizes (128, 192, and 256 bits). The SU is controlled via a custom instruction set extension, enabling faster multi-mode AES operations compared to standard RISC-V instructions.	To ensure high-performance operation of the SU, a high-bandwidth buffer system is implemented for intermediate data exchange between the SU and DM.
	
	\subsection{High-Bandwidth Internal Buffers for Fast Continuous Data Accessing}

	\begin{figure}[t]
		\centering
		\includegraphics[width=0.47 \textwidth]{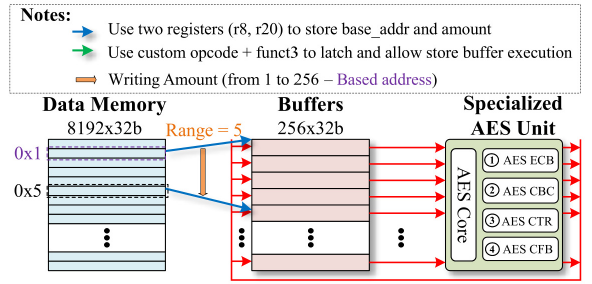}
		\vspace{-2mm}
`		\caption{High-bandwidth Buffer Set for Fast Continuous Data Accessing}
		\vspace{-2mm}
		\label{fig:buffer}
	\end{figure}

        \begin{figure}[t]
		\centering
		\includegraphics[width=0.47 \textwidth]{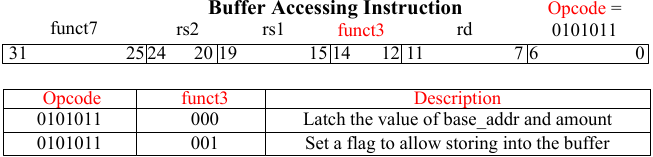}
		\vspace{-2mm}
		\caption{Structure of the Buffer Accessing Instructions.}
		\vspace{-2mm}
		\label{fig:buffer_instruction}
	\end{figure}

	The use of the ALU in the RISC-V core, following the traditional RISC-V resource usage rules, requires many instructions to organize the data needed for primary processing. As a consequence, the AES computation performance of the conventional RISC-V core is extremely low since it does not meet the current security requirements. Therefore, the proposed AES-RV is improved by using a SAU that contains specialized components optimized to handle multi-mode AES. Accordingly, the SAU block requires substantial data for continuous computations through the computation loops. To ensure that the computation efficiency is not affected, a set of 256 32-bit buffers is placed before the SAU to hold large amounts of data such as the key, initial vector (IV), plain text, and cipher text. This buffer set can load and store large amounts of data in just one cycle, allowing the SAU to compute continuously without data interruption.

	Fig. \ref{fig:buffer} shows the detailed communication mechanism of the high-bandwidth buffer set with the data memory. In this scheme, a quantity of data can be transferred between the data memory (DM) and the buffer based on the management of the values in registers r8 and r20. Register r8 stores the DM address for reading or writing, while register r20 contains the number of 32-bit values to be read or written. This process is executed by the buffer accessing instruction described in Fig. \ref{fig:buffer_instruction}. The buffer consists of one instruction that specifies the address and amount and another instruction to set the flag to start the read/write process. Once the buffer set is fully loaded with data, the SAU can load the entire content and immediately perform the AES Key Expansion and main round computation. When using the original RISC-V instruction set, after one round of computation, the system has to execute many instructions to rearrange the data and store it in the BRAM. By using the high-bandwidth buffer set, this process is completely eliminated. It should be noted that each data rearrangement requires many instructions and takes even longer than the AES computation process.

	\subsection{Specialized AES Unit (SAU) with Low-Latency Custom Instruction Extension}

	\begin{figure}[t]
		\centering
		\includegraphics[width=0.47 \textwidth]{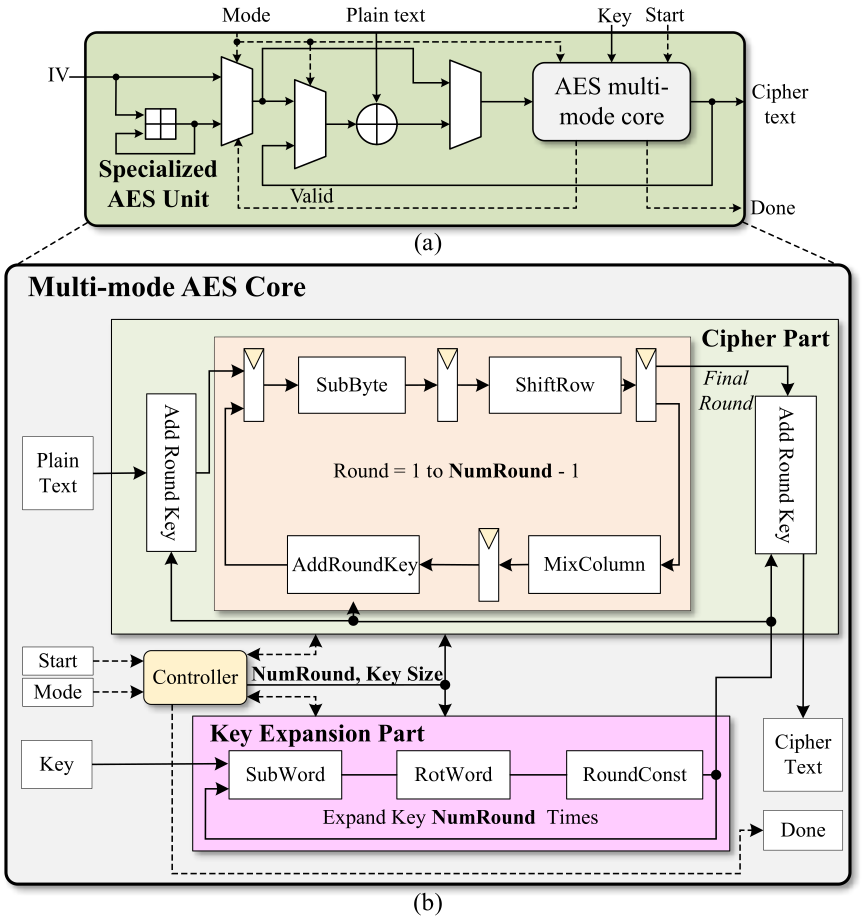}
		\vspace{-2mm}
		\caption{Specialized AES Unit and Multi-mode AES Core architecture.}
		\vspace{-2mm}
		\label{fig:alu}
	\end{figure}
	
	The use of conventional RISC-V instructions to implement AES results in high latency due to the large number of instructions required. To address this issue, we propose a Specialized AES Unit (SAU) as shown in Fig.~\ref{fig:alu} to accelerate AES computation. The SAU processes data including plain text, key, IV, and cipher text. Its control signals consist of \texttt{start} and \texttt{done}, while the configuration data, which contains the operating mode, determines the number of rounds and key size for the AES multi-mode core. The multi-mode AES core comprises three main components: the cipher part, the key expansion part, and the controller. The cipher part executes the main AES computation loop through four steps: AddRoundKey, SubByte, ShiftRow, and MixColumn. The number of computation rounds and the AES key size depend on the input mode, as determined by the controller. To minimize the critical path, the cipher part employs a 4-stage pipeline architecture. The key expansion part is responsible for generating the necessary round keys and consists of three main functions: SubWord, RotWord, and RoundConst. These functions expand the original key for use in the cipher part’s computation rounds.
	
\begin{figure}[t]
		\centering
		\includegraphics[width=0.47 \textwidth]{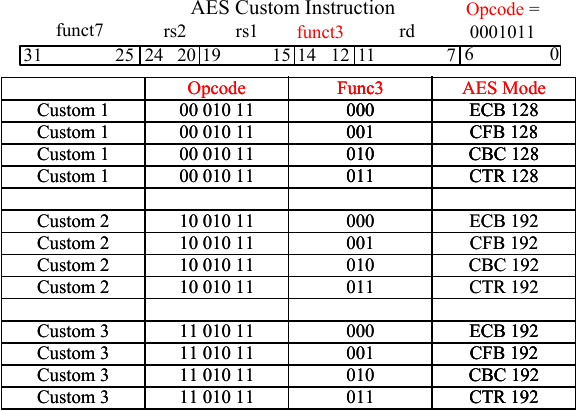}
		\vspace{-2mm}
		\caption{Custom instruction for calling Specialized AES Unit.}
		\vspace{-2mm}
		\label{fig:instr_alu}
	\end{figure}

	Control of the SAU is achieved through AES custom instructions, as illustrated in Fig.~\ref{fig:instr_alu}. Three groups of custom instructions are defined by opcodes corresponding to key sizes of 128, 192, and 256 bits. By altering the \texttt{func3} field, the SAU supports up to four AES modes: ECB, CFB, CBC, and CTR. In summary, the integration of the SAU and the internal buffer set significantly enhances the performance of multi-mode AES computation.

    	\begin{figure}[t]
		\centering
		\includegraphics[width=0.47 \textwidth]{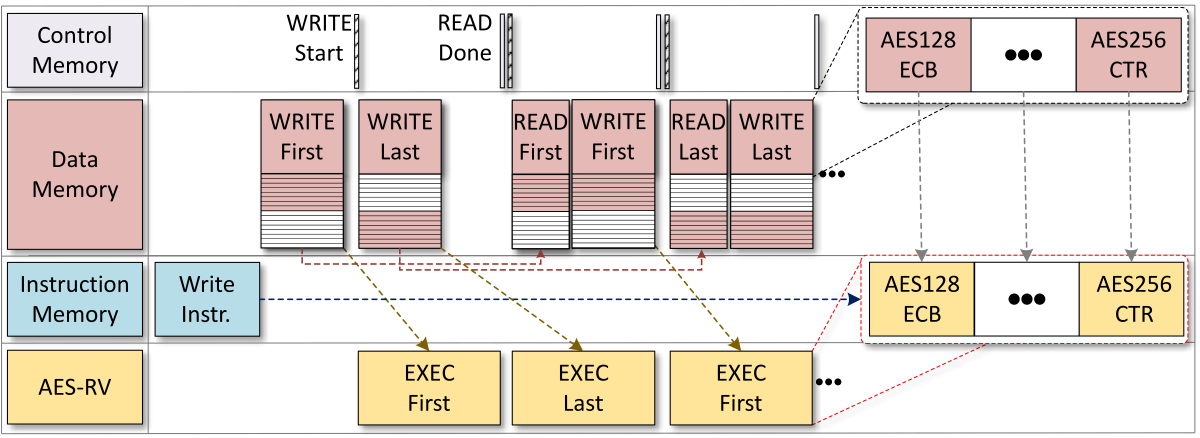}
		\vspace{-2mm}
		\caption{Timing diagram of the system pipeline using ping-pong memory transfer mechanism.}
		\vspace{-2mm}
		\label{fig:system_pipeline}
	\end{figure}
	
	\subsection{System Pipelining with a Ping-Pong Memory Transfer Mechanism}
		\begin{figure*}[t]
		\centering
            \includegraphics[width=0.98 \textwidth]{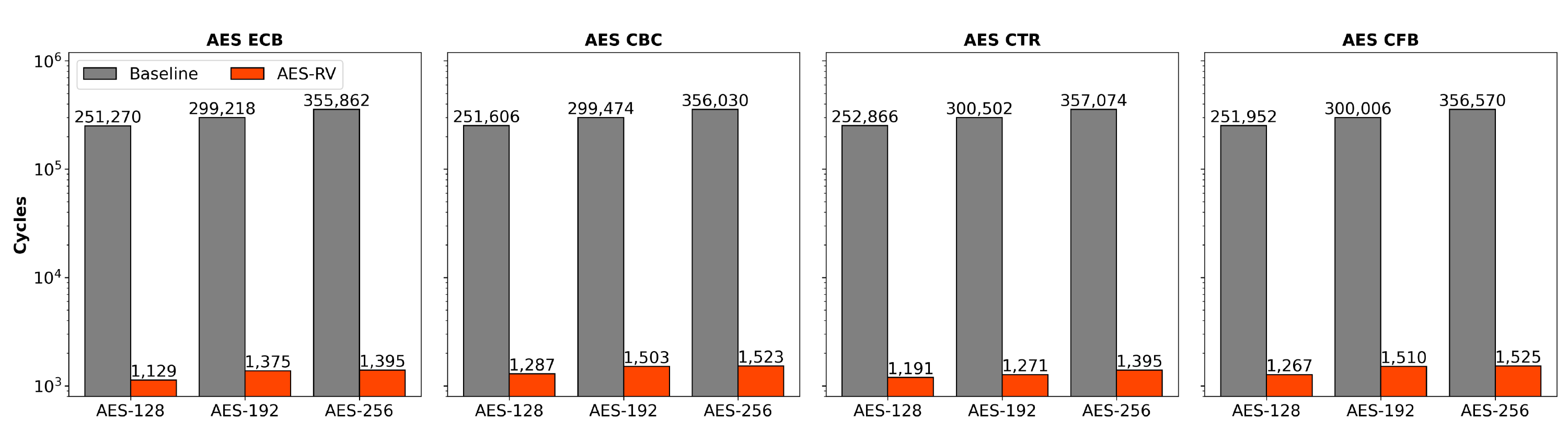}
		\vspace{-2mm} 
		\caption{Performance comparison of AES-RV and baseline RISC-V implementation in terms of execution cycles across different AES modes and key sizes.}
		\vspace{-2mm}
		\label{fig:baseline}
	\end{figure*}
	
	In practice, transferring a large amount of data to the AES-RV core via the AXI interface introduces significant latency. In a straightforward system, the waiting time for reading and writing computational data can exceed the hardware processing time, leading to a system bottleneck. In other words, merely accelerating the AES-RV core cannot substantially improve the overall system performance. To address this issue, a system pipeline with a ping-pong memory transfer mechanism is proposed to mitigate the bottleneck caused by data transfer.
	
	Fig.~\ref{fig:system_pipeline} illustrates the timing schedule of the proposed system pipeline with the ping-pong memory transfer mechanism. The data memory is divided into two equal parts: \textit{first} and \textit{last}. Initially, instructions are loaded into the AES-RV core with configurations that support multiple AES modes and key sizes. Subsequently, the input data is filled into the first half of the data memory (WRITE First), followed by a control signal to initiate the AES-RV core for processing this portion (EXEC First). During the EXEC First phase, new data is written into the second half of the data memory (WRITE Last). Notably, if the AES-RV core completes the EXEC First phase, the system performs multiple tasks during the next execution cycle. It simultaneously reads the processed data from the first half of the data memory (referred to as READ First). At the same time, it writes new input data into that same memory region (WRITE First). Meanwhile, the AES-RV core processes the second half of the data memory (EXEC Last). This approach allows the processing system to read the output and write new input data into the alternate memory region while the AES-RV core continues computation. As a result, the entire SoC system eliminates idle wait times for data transfer.
	
	For optimal performance, data transfer time must be shorter than hardware execution time, ensuring that system performance aligns with hardware computing speed, regardless of transfer latency.

	\section{Evaluation Results}
	\label{sec3}
	
	\subsection{Implementation results on ZCU102 FPGA SoC}

        To evaluate the correctness and practicality, AES-RV was implemented on the Zynq UltraScale+ MPSoC ZCU102 FPGA SoC, as illustrated in Fig.~\ref{fig:overview}. The processing system is managed by an ARM Cortex-A53 CPU running Debian GNU/Linux 11, installed via PetaLinux 2022.2. The programmable logic hosts the AES-RV IP, synthesized using Vivado Design Suite 2022.2. AES-RV was verified by executing all AES functions with key sizes from the set $\mathcal{K} = \{128, 192, 256\} \text{ bits}$ across multiple modes defined by the set $\mathcal{M} = \{\text{ECB, CBC, CTR, CFB}\}$. The real-time SoC verification demonstrated that AES-RV successfully processed 100,000 random plaintext inputs for each mode with 100\% accuracy at a frequency of 200 MHz. Utilization reports indicate that AES-RV occupies 29,608 flip-flops (FFs), 32,483 lookup tables (LUTs), and 12 Block RAM tiles (36 KB each). Furthermore, power analysis shows that AES-RV consumes a total power of 4.043 W, with the AES-RV IP contributing 0.043 W in dynamic power.

        In general, AES-RV exhibits the ability to operate at high frequencies on real-time SoC systems while consuming minimal power, rendering it an ideal candidate for integration into SoC-based applications.

	\subsection{Performance Comparison of AES-RV and Baseline RISC-V Implementation}
    
        In this section, the execution cycles of AES-RV when computing AES for all key sizes $\mathcal{K}$ and modes $\mathcal{M}$ on four consecutive data blocks are compared in detail with the baseline RISC-V implementation. The comparison results are obtained through waveform simulation extraction, as detailed in Fig.~\ref{fig:baseline}. 
        
        For the AES-ECB mode, AES-RV achieves execution cycles ranging from 1,129 to 1,395 cycles, outperforming the baseline RISC-V by \textbf{217.61 times} (251,270 cycles vs. 1,129 cycles) to \textbf{255.10 times} (355,862 cycles vs. 1,395 cycles). In the AES-CBC mode, AES-RV completes encryption within 1,287 to 1,523 cycles, demonstrating a speedup of \textbf{195.50 times} (251,606 cycles vs. 1,287 cycles) to \textbf{233.77 times} (356,030 cycles vs. 1,523 cycles) compared to the baseline. For the AES-CTR mode, AES-RV executes in 1,191 to 1,395 cycles, surpassing the baseline RISC-V with a performance gain from \textbf{212.31 times} (252,866 cycles vs. 1,191 cycles) to \textbf{255.97 times} (357,074 cycles vs. 1,395 cycles). Finally, in the AES-CFB mode, AES-RV achieves 1,267 to 1,525 cycles, significantly outperforming the baseline by \textbf{198.68 times} (251,952 cycles vs. 1,267 cycles) to \textbf{233.82 times} (356,570 cycles vs. 1,525 cycles). 
        
        Generally, AES-RV significantly outperforms the baseline RISC-V by leveraging custom instructions and buffer optimizations.
    	\renewcommand{\arraystretch}{1.3} 
	
	\begin{table*}[h]
		\centering
		\resizebox{\textwidth}{!}{ 
			\begin{threeparttable}
				\caption{Comparison of AES Implementations on hardware platforms in terms of Throughput and Hardware Efficiency.}
				\label{tab:aes_comparison}
				\begin{tabular}{|c|c|c|c|c|c|c|c|c|c|c|}
					\hline
					\textbf{References} & \textbf{Devices} & \makecell{\textbf{$\text{F}_\text{max}$} \\ \textbf{(MHz)}} & \textbf{FFs} & \textbf{LUTs} & \textbf{BRAMs} & \#\textbf{Slices}$^{\dagger\dagger}$ & \makecell{\textbf{Power} \\ \textbf{(W)}} & \makecell{\textbf{Throughput} \\ \textbf{(Mbps)}} & \makecell{\textbf{Energy} \\ \textbf{Efficiency} \\ \textbf{(Mbps/W)}} & \makecell{\textbf{Area} \\ \textbf{Efficiency} \\ \textbf{(Mbps/Slice)}} \\
					\hline
					\multirow{2}{*}{ATC 2024 \cite{le2024rvcp}} & \multirow{2}{*}{ZCU102 FPGA} & \multirow{2}{*}{210} & 7,584$^*$ & 7,562$^*$ & \multirow{2}{*}{16} & 2,531$^*$ & 0.136$^*$ & \multirow{2}{*}{4.81} & 3.54E+01$^*$ & 1.90E-04$^*$ \\
					& & & 29,644$^\dagger$ & 34,898$^\dagger$ & & 9,365$^\dagger$ & 4.22$^\dagger$ & & 1.14E+00$^\dagger$ & 5.14E-04$^\dagger$ \\
					\hline
					\multirow{1}{*}{ICECS 2024 \cite{10848921}} & 22nm FDSOI ASIC & 1,000 & - & - & - & - & 0.008 & 7.07 & 8.84E+02 & - \\
					\hline
					\multirow{1}{*}{PRIME 2024 \cite{9547962}} & Nexys Artix-7 FPGA & 50 & - & 609 & - & 33,650 & 397 & 2.86 & 1.15E+02 & 8.50E-05 \\
					\hline
					\multirow{1}{*}{DDECS 2024 \cite{10508919}} & PYNQ Z2 FPGA & 100 & 10,454 & 15,885 & - & 7,943 & - & 4.72 & - & 5.94E-04 \\
					\hline
					\multirow{2}{*}{AES-RV} & \multirow{2}{*}{ZCU102 FPGA} & \multirow{2}{*}{241} & 7,548$^*$ & 5,147$^*$ & \multirow{2}{*}{12} & 1,767$^*$ & 0.046$^*$ & \multirow{2}{*}{95.88} & 2.08E+03$^*$ & 5.43E-02$^*$ \\
					& & & 29,608$^\dagger$ & 32,483$^\dagger$ & & 8,601$^\dagger$ & 4.043$^\dagger$ & & 2.37E+01$^\dagger$ & 1.11E-02$^\dagger$ \\
					\hline
				\end{tabular}
				
				\begin{tablenotes}
                    \item[] $^{(*)}$ Denotes results for the AES core only; $^{(\dagger)}$ Denotes results for the entire SoC.
                    \item $^{(\dagger\dagger)}$ PYNQ Z2: 1 Slice = 6 LUTs, 8 FFs; ZCU102: 1 Slice = 4 LUTs, 8 FFs, and 1 BRAM 36Kb = 40 slices.
                \end{tablenotes}
			\end{threeparttable}}
	\end{table*}
    
	\subsection{Performance and Energy Efficiency Evaluation on Real-Time Software Platforms}
		\begin{figure}[t]
		\centering
		\includegraphics[width=0.47 \textwidth]{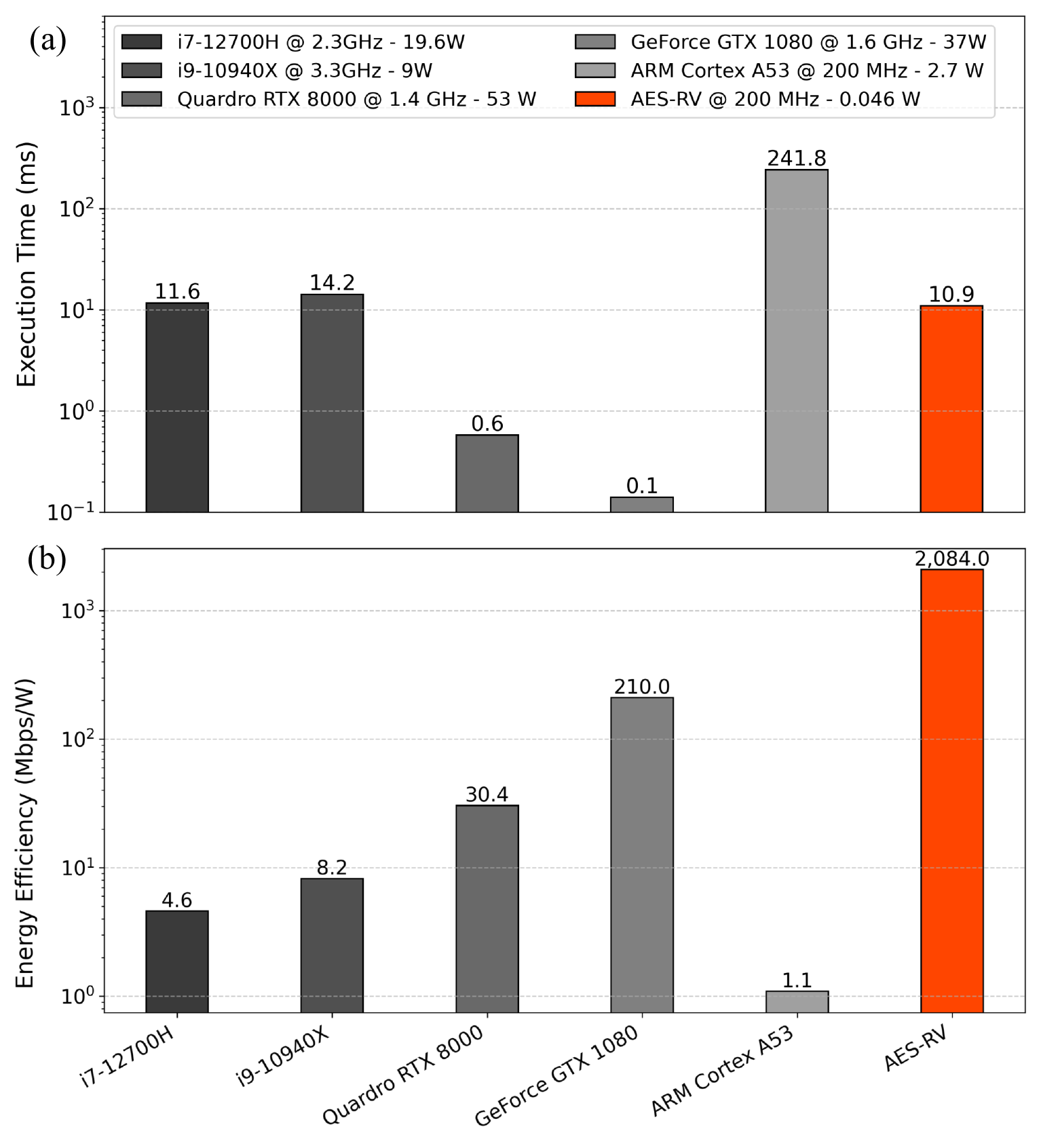}
		\vspace{-2mm}
		\caption{Comparisons with powerful CPUs/GPUs on (a) execution time and (b) energy efficiency.}
		\vspace{-2mm}
		\label{fig:software_compare}
	\end{figure}

        To assess performance and energy efficiency, the proposed hardware was benchmarked on 8,192 AES-CBC test cases (128-bit key) and compared against high-performance CPUs (Intel Core i7-12700H, Core i9-10940X), GPUs (Quadro RTX 8000, GeForce GTX 1080 with 2048-thread execution), and the ARM Cortex A53 on the ZCU102 FPGA SoC for energy efficiency. Detailed results are shown in Fig.~\ref{fig:software_compare}.

        In the \textbf{execution time comparison} shown in Fig.~\ref{fig:software_compare} (a), AES-RV demonstrates comparable performance to \textbf{Intel Core i7-12700H @ 2.3GHz}, achieving a speedup of \textbf{1.06 times} (11.6 ms vs. 10.9 ms). Furthermore, AES-RV outperforms Intel Core i9-10940X @ 3.3GHz by \textbf{1.30 times} (14.2 ms vs. 10.9 ms). Compared to ARM Cortex A53 CPUs @ 200MHz, AES-RV exhibits a remarkable speedup of \textbf{22.18 times} (241.8 ms vs. 10.9 ms).  Regarding high-performance GPUs such as Quadro RTX 8000 @ 1.4 GHz and GeForce GTX 1080 @ 1.6 GHz, AES-RV does not surpass them in execution time, as GPUs complete AES operations in 0.6 ms and 0.1 ms, respectively. However, these CPUs and GPUs operate at significantly higher power consumption, resulting in inferior energy efficiency. The energy efficiency comparison, measured in Mbps/W and detailed in Fig.~\ref{fig:software_compare} (b), highlights the substantial advantage of AES-RV over traditional CPUs and GPUs. Specifically, AES-RV achieves \textbf{453.04 times} (4.6 Mbps/W vs. 2084 Mbps/W), higher efficiency than Intel Core i7-12700H @ 19.6W  \textbf{254.15 times} (8.2 Mbps/W vs. 2084 Mbps/W) higher efficiency than Intel Core i9-10940X @ 9W , and \textbf{1894.55 times} (1.1 Mbps/W vs. 2084 Mbps/W) higher efficiency than ARM Cortex A53 @ 2.7W . Similarly, AES-RV also outperforms high-power GPUs, providing \textbf{68.55 times} (30.4 Mbps/W vs. 2084 Mbps/W) better energy efficiency than Quadro RTX 8000 @ 53W  and \textbf{9.92 times} (210 Mbps/W vs. 2084 Mbps/W) better energy efficiency than GeForce GTX 1080 @ 37W.
        
        Overall, AES-RV matches modern CPUs in speed while vastly outperforming CPUs and GPUs in energy efficiency, making it ideal for low-power, real-time applications.

	\subsection{Comprehensive Performance and Hardware Efficiency Evaluation for AES Implementations}
	
        To demonstrate the improvements in hardware efficiency, key evaluation metrics, including maximum frequency, hardware resource utilization, power consumption, throughput, energy efficiency, and area efficiency, are summarized in Table~\ref{tab:aes_comparison} for comparison with related work \cite{le2024rvcp, 10848921, 9547962, 10508919}. Notably, only AES-RV and \cite{le2024rvcp} support full SoC execution, whereas other implementations focus solely on core-level development. The formulas for throughput, energy efficiency, and area efficiency are defined in Eq.~(1)–(3).

        \begin{align}
            \text{Throughput} &= \frac{\text{F}_\text{max} \times \text{Block Size} }{\text{Cycles per Block}} \quad \text{(Mbps)} \\
            \text{Energy Efficiency} &= \frac{\text{Throughput}}{\text{Power Consumption}} \quad \text{(Mbps/W)} \\
            \text{Area Efficiency} &= \frac{\text{Throughput}}{\text{Total Slices}} \quad \text{(Mbps/Slice)}
        \end{align}
        
        Compared to \cite{le2024rvcp} in ATC 2024, AES-RV supports a maximum operating frequency that is \textbf{1.15 times} higher (241 MHz vs. 210 MHz), while also utilizing fewer hardware resources for both the core and SoC by \textbf{1.43/1.09 times} (1,767/8,601 vs. 2,531/9,365 slices). Additionally, it consumes \textbf{2.96/1.05 times} less power (0.046/4.043 W vs. 0.136/4.22 W). In terms of performance, AES-RV achieves \textbf{19.94 times} higher throughput (95.88 Mbps vs. 4.81 Mbps). Moreover, AES-RV surpasses \cite{le2024rvcp} in energy efficiency by \textbf{58.76/20.79 times} (2084.0/23.7 Mbps/W vs. 35.4/1.14 Mbps/W) and in area efficiency by \textbf{285.26/21.6 times} (0.0543/0.0111 Mbps/Slice vs. 0.00019/0.00051 Mbps/Slice). Similarly, compared to \cite{10848921} in ICECS 2024, AES-RV delivers \textbf{13.56 times} higher throughput (95.88 Mbps vs. 7.07 Mbps) and achieves \textbf{2.36 times} greater energy efficiency (2084.0 Mbps/W vs. 884 Mbps/W). When comparing with \cite{9547962} in PRIME 2024, AES-RV operates at a maximum frequency that is \textbf{4.82 times} higher (241 MHz vs. 50 MHz), while also utilizing \textbf{19.05 times} fewer core resources (1,767 slices vs. 33,650 slices) and consuming \textbf{8,630 times} less power (0.046 W vs. 397 W). In terms of performance, AES-RV achieves \textbf{33.52 times} higher throughput (95.88 Mbps vs. 2.86 Mbps) and further demonstrates \textbf{18.08 times} greater energy efficiency (2084.0 Mbps/W vs. 115.2 Mbps/W), along with \textbf{638.8 times} better area efficiency (0.0543 Mbps/Slice vs. 0.000085 Mbps/Slice). Finally, compared to \cite{10508919} in DDECS 2024, AES-RV supports a maximum frequency \textbf{2.41 times} higher (241 MHz vs. 100 MHz) while requiring \textbf{4.5 times} fewer core resources (1,767 slices vs. 7,943 slices). Regarding performance, AES-RV achieves \textbf{20.32 times} higher throughput (95.88 Mbps vs. 4.72 Mbps) and outperforms in area efficiency by \textbf{91.42 times} (0.0543 Mbps/Slice vs. 0.000594 Mbps/Slice).
        
        In conclusion, AES-RV demonstrates superior throughput and hardware efficiency, benefiting from optimized instruction sets and an efficient architecture tailored for real-time SoC environments.

\section{Conclusion}
    \label{sec4}
    This paper proposes a hardware-efficient RISC-V accelerator with low-latency AES instruction extension (AES-RV), enhancing flexibility and energy efficiency in cryptographic applications. By integrating three key optimizations—high-bandwidth internal buffers, a specialized AES unit with low-latency custom instructions, and system pipelining with a ping-pong memory mechanism—AES-RV significantly improves processing speed and hardware efficiency over conventional AES implementations. FPGA SoC experiments demonstrate notable gains in performance and energy efficiency, making AES-RV a strong candidate for secure, high-performance embedded systems. Future work will extend AES-RV with additional instruction sets for post-quantum cryptography, including CRYSTALS-Kyber and CRYSTALS-Dilithium, further broadening its applicability.



\bibliographystyle{ieeetr}
\bibliography{references.bib}

\begin{thebibliography}{10}

\bibitem{sharma2022survey}
M.~Sharma and et~al., ``{A Survey of RISC-V CPU for IoT Applications},'' in
  {\em Proceedings of the International Conference on Innovative Computing \&
  Communication (ICICC) 2022}, SSRN, February 2022.

\bibitem{a1}
J.~Park and et~al., ``{Designing Low-Power RISC-V Multicore Processors With a
  Shared Lightweight Floating Point Unit for IoT Endnodes},'' {\em IEEE
  Transactions on Circuits and Systems I: Regular Papers}, vol.~71,
  pp.~4106--4119, 2024.

\bibitem{a2}
T.-T. Hoang and et~al., ``{Low-power high-performance 32-bit RISC-V
  microcontroller on 65-nm silicon-on-thin-BOX (SOTB)},'' {\em IEICE
  Electronics Express}, vol.~17, pp.~20200282--20200282, 2020.

\bibitem{a3}
K.-D. Nguyen and et~al., ``{A trigonometric hardware acceleration in 32-bit
  RISC-V microcontroller with custom instruction},'' {\em IEICE Electronics
  Express}, vol.~18, no.~16, pp.~20210266--20210266, 2021.

\bibitem{a4}
M.~Liu, ``{A co-design method of customized ISA design space exploration and
  fixed-point library construction for RISC-V dedicated processor},'' {\em
  IEICE Electronics Express}, vol.~19, no.~13, pp.~20220244--20220244, 2022.

\bibitem{a5}
Q.~Yin and et~al., ``{Design and implementation of RISC-V system-on-chip for
  SPWM generation based on FPGA},'' {\em IEICE Electronics Express}, vol.~21,
  no.~24, pp.~20240603--20240603, 2024.

\bibitem{Le2024RISCVCoprocessor}
D.~H.~A. Le and et~al., ``{High-Efficiency {RISC-V}-Based Cryptographic
  Coprocessor for Security Applications},'' in {\em International SoC Design
  Conference ({ISOCC})}, pp.~103--104, 2024.

\bibitem{a6}
Q.~Zou and et~al., ``{28nm asynchronous area-saving AES processor with high
  Common and Machine learning side-channel attack resistance},'' {\em IEICE
  Electron. Express}, vol.~18, p.~20210309, 2021.

\bibitem{a7}
Leurent and et~al., ``{New representations of the AES key schedule},'' {\em J.
  Cryptol.}, vol.~38, p.~1, 2025.

\bibitem{a8}
W.~K. Lee and et~al., ``{Efficient Implementation of AES-CTR and AES-ECB on
  GPUs With Applications for High-Speed FrodoKEM and Exhaustive Key Search},''
  {\em IEEE Trans. Circuits Syst. II Express Briefs}, vol.~69, p.~2962, 2022.

\bibitem{a9}
Y.~Zhang and et~al., ``{A lightweight AES algorithm implementation for
  encrypting voice messages using field programmable gate arrays},'' {\em J.
  King Saud Univ. - Comput. Inf. Sci.}, vol.~34, p.~3878, 2022.

\bibitem{a10}
N.~H. Nguyen, , and et~al., ``{LI-RV: A Fast and Efficient RISC-V based
  Coprocessor for Lightweight Cryptography},'' in {\em 2024 21st International
  SoC Design Conference (ISOCC)}, pp.~1--2, 2024.

\bibitem{a18}
K.~Stangherlin and M.~Sachdev, ``{Design and Implementation of a Secure RISC-V
  Microprocessor},'' {\em IEEE Trans. Very Large Scale Integr. VLSI Syst.},
  vol.~30, p.~1705, 2022.

\bibitem{a19}
Pan and et~al., ``{A Lightweight AES Coprocessor Based on RISC-V Custom
  Instructions},'' {\em Secur. Commun. Netw.}, p.~9355123, 2021.

\bibitem{a20}
O.~Simola and et~al., ``{RISC-V Core with AES-256 Accelerator},'' in {\em 31st
  IEEE International Conference on Electronics, Circuits and Systems (ICECS)},
  p.~1, 2024.

\bibitem{a21}
Ignatius and et~al., ``{Power Side-Channel Attacks on Crypto-core based on
  RISC-V ISA for High-security Applications},'' in {\em IEEE Access}, vol.~12,
  p.~150230, 2024.

\bibitem{a22}
Cheng and et~al., ``{A Hardware Security Evaluation Platform on RISC-V SoC},''
  in {\em IEEE International Test Conference in Asia (ITC-Asia)}, p.~1, 2024.

\bibitem{a23}
O.~Simola and et~al., ``{RISC-V Core with AES-256 Accelerator},'' in {\em 31st
  IEEE International Conference on Electronics, Circuits and Systems (ICECS)},
  p.~1, 2024.

\bibitem{9759828}
Salman and et~al., ``{Lightweight Modifications in the Advanced Encryption
  Standard (AES) for IoT Applications: A Comparative Survey},'' in {\em 2022
  International Conference on Computer Science and Software Engineering
  (CSASE)}, pp.~325--330, 2022.

\bibitem{a11}
S.~Jeon and et~al., ``{Cross-Layer Encryption of CFB-AES-TURBO for Advanced
  Satellite Data Transmission Security},'' {\em IEEE Trans. Aerosp. Electron.
  Syst.}, vol.~58, p.~1, 2022.

\bibitem{a12}
E.~Choi and et~al., ``{AESware: Developing AES-enabled low-power multicore
  processors leveraging open RISC-V cores with a shared lightweight AES
  accelerator},'' {\em Eng. Sci. Technol. Int. J.}, vol.~60, p.~1, 2024.

\bibitem{a13}
C.~Duran and E.~Roa, ``{A 10pJ/bit 256b AES-SoC Exploiting Memory Access
  Acceleration},'' {\em IEEE Trans. Circuits Syst. II Express Briefs}, vol.~69,
  p.~1612, 2022.

\bibitem{a14}
Y.~M. Kuo and et~al., ``{RISC-V Galois Field ISA Extension for Non-Binary
  Error-Correction Codes and Classical and Post-Quantum Cryptography},'' {\em
  IEEE Trans. Comput.}, vol.~72, p.~682, 2023.

\bibitem{a15}
P.~Nannipieri and et~al., ``{VLSI Design of Advanced-Features AES
  Cryptoprocessor in the Framework of the European Processor Initiative},''
  {\em IEEE Trans. Very Large Scale Integr. VLSI Syst.}, vol.~30, p.~177, 2022.

\bibitem{a16}
W.~Wang and et~al., ``{An energy-efficient crypto-extension design for
  RISC-V},'' {\em Microelectron. J.}, vol.~115, p.~105165, 2021.

\bibitem{a17}
D.~Reis and et~al., ``{IMCRYPTO: An In-Memory Computing Fabric for AES
  Encryption and Decryption},'' {\em IEEE Trans. Very Large Scale Integr. VLSI
  Syst.}, vol.~30, p.~553, 2022.

\bibitem{le2024rvcp}
V.~T.~D. Le and et~al., ``{RVCP: High-Efficiency RISC-V Co-Processor for
  Security Applications in IoT and Server Systems},'' in {\em 2024
  International Conference on Advanced Technologies for Communications (ATC)},
  IEEE, 2024.

\bibitem{10848921}
Simola and et~al., ``{RISC-V Core with AES-256 Accelerator},'' in {\em 2024
  31st IEEE International Conference on Electronics, Circuits and Systems
  (ICECS)}, pp.~1--4, 2024.

\bibitem{9547962}
Zgheib and et~al., ``{Extending a RISC-V core with an AES hardware accelerator
  to meet IOT constraints},'' in {\em SMACD / PRIME 2021; International
  Conference on SMACD and 16th Conference on PRIME}, pp.~1--4, 2021.

\bibitem{10508919}
M.~N. Rizi and et~al., ``{Optimised AES with RISC-V Vector Extensions},'' in
  {\em 2024 27th International Symposium on Design and Diagnostics of
  Electronic Circuits and Systems (DDECS)}, pp.~57--60, 2024.

\end{thebibliography}



\end{document}